\documentclass[journal,letterpaper,twocolumn]{IEEEtran}
\usepackage{cite}
\usepackage{fancyhdr}
\usepackage{float}
\usepackage{graphicx}
\usepackage{epsfig}
\usepackage{bm}
\usepackage{amsmath}
\usepackage{mathrsfs}
\usepackage{subfigure}
\usepackage{array}
\usepackage{booktabs}
\usepackage{amssymb}
\usepackage{graphicx}
\usepackage{epstopdf}
\usepackage{algorithm}
\usepackage{amsthm}
\usepackage{bbm}
\usepackage{graphicx}
\usepackage{epsfig}
\usepackage{color}

\makeatother

\title{On Safeguarding Privacy and Security in the Framework of Federated Learning}
\begin{document}
\author{\IEEEauthorblockN{Chuan Ma\thanks{This work is supported in part by the National Key R$\&$D Program under Grants 2018YFB1004800, and by National Natural Science Foundation of China under 61872184 and 61727802, and by the U.S. National Science Foundation under Grant CCF-1908308. The corresponding authors are Jun Li and Feng Shu.}
\thanks{C. Ma, J. Li and F. Shu are with the School of Electronic and Optical Engineering, Nanjing University of Science and Technology, Nanjing, China. (e-mail: \{chuan.ma, jun.li, feng.shu\}@njust.edu.cn).},
Jun Li, \emph{Senior Member, IEEE},
\thanks{J. Li is also with the Department of Software Engineering, Institute of Cybernetics, National Research Tomsk Polytechnic University, Tomsk, 634050, Russia.}
Ming Ding, \emph{Senior Member, IEEE},
\thanks{M. Ding is with Data61, CSIRO, Australia (e-mail: Ming.Ding@data61.csiro.au).}\\
Howard H. Yang, \emph{Member, IEEE},
\thanks{H. H. Yang and T. Q. S. Quek are with the Information System Technology and Design Pillar, Singapore University of Technology and Design (e-mail: \{howard yang, tonyquek\}@sutd.edu.sg).}
Feng Shu, \emph{Member, IEEE}\\
Tony Q. S. Quek, \emph{Fellow, IEEE},
and H. Vincent Poor, \emph{Fellow, IEEE}
\thanks{H. V. Poor is with the Department of Electrical Engineering, Princeton University, Princeton, NJ 08544 USA (e-mail: poor@princeton.edu).}
}}

\maketitle
\begin{abstract}
Motivated by the advancing computational capacity of wireless end-user equipment (UE), as well as the increasing concerns about sharing private data, a new machine learning (ML) paradigm has emerged, namely federated learning (FL). Specifically, FL allows a decoupling of data provision at UEs and ML model aggregation at a central unit. By training model locally, FL is capable of avoiding direct data leakage from the UEs, thereby preserving privacy and security to some extend. However, even if raw data are not disclosed from UEs, individual's private information can still be extracted by some recently discovered attacks against the FL architecture. In this work, we analyze the privacy and security issues in FL, and discuss several challenges on preserving privacy and security when designing FL systems. In addition, we provide extensive simulation results to showcase the discussed issues and possible solutions.
\end{abstract}


\section{Introduction}

Recent technological advancements are currently transforming the ways in which data is created and processed. With the advent of the internet-of-things (IoT), the number of intelligent devices in the world is rapidly growing in the last couple of years. Many of these devices are equipped with various sensors and increasingly powerful hardware, which allow them to not just collect, but more importantly, process data at unprecedented scales.
In a concurrent development, artificial intelligence (AI) has revolutionized the ways that information is extracted with ground breaking successes in areas such as computer vision, natural language processing, voice recognition, etc\cite{wang2018edge}. Therefore, there is high demand for harnessing the rich data provided by distributed devices to improve machine learning models.

At the same time, data privacy has become a growing concern for clients.  In particular, the emergence of centralized searchable data repositories has made the leakage of private information, e.g. health conditions, travel information, and financial data, an urgent social problem \cite{liu2019adversaries}. Furthermore, the diverse set of open data applications, such as census data dissemination and social networks, place more emphasis on privacy concerns. In such practices, the access to real-life datasets may cause information leakage even in pure research activities. Consequently, privacy preservation has become a critical issue.

To tackle the challenge of protecting individuals' privacy, a new paradigm has emerged, i.e., federated learning (FL) \cite{konevcny2016federated}, which allows a decoupling of data provision at end-user equipment (UE) and machine learning model aggregation, such as network parameters of deep learning, at a centralized server. The purpose of FL is to cooperatively learn a global model without sacrificing the data privacy directly.
In particular, FL has distinct privacy advantages compared to data center training on a {\color{black}dataset}. At a server, holding even an ``anonymized" dataset can still put client privacy at risk via linkage to other datasets. 
In contrast, the information transmitted for FL consists of the minimal updates to improve a particular machine learning model. The updates themselves can be ephemeral, and will never contain more information than the raw training data (by the data processing inequality). Further, the source of the updates is not needed by the aggregation algorithm, so updates can be transmitted without identifying metadata over a mixed network such as Tor \cite{chaum1981untraceable} or via a trusted third party. These generic approaches include de-identification methods like anonymization \cite{machanavajjhala2006diversity}, obfuscation methods like differential privacy \cite{dwork2006calibrating}, cryptographic techniques like homomorphic encryption \cite{papernot2016semi} and secure multi-party computation (SMC) protocols like oblivious transfer and garbled circuits \cite{rosulek2017improvements}.

However, although the data is not explicitly shared in the original format, it is still possible for adversaries to reconstruct the raw data approximately,
especially when the architecture and parameters are not completely protected. In addition, FL can expose intermediate results such as parameter updates from an optimization algorithm like stochastic gradient descent (SGD), and the transmission of these gradients may actually leak private information \cite{8241854} when exposed together with a data structure such as image pixels. In addition, the existence of malevolent users may induce further security issues. Therefore, the design of FL still needs further protection of parameters as well as investigations on the tradeoffs between the privacy-security-level and the system performance.

Inspired by this research gap, we briefly investigate the potential privacy and security issues in FL. Specifically, we clarify that the current protection methods are mainly focused on the server and client side, and then investigate four important aspects of current designs, including convergence, data poisoning, scaling up and model aggregation.
The remainder of this article is organized as follows. Section II introduces the basic model and key directions on the protection of FL. Section III illustrates challenges and opportunities in developing private and secure FL, and Section IV provides probable solutions and future work for discussion. Finally, conclusions are drawn in Section V.

\section{Background}

\begin{figure}
\centering
  \includegraphics[width=0.44\textwidth]{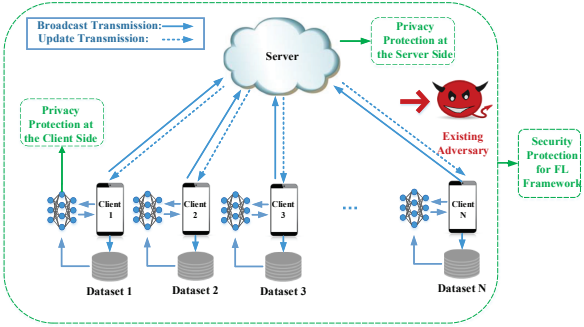}\\
\caption{The structure of private and secure federated learning framework}\label{system}
\end{figure}

We first introduce the basic model of FL, which is illustrated in Fig.~\ref{system}. As can be seen from Fig.~\ref{system}, each client downloads a globally shared model from the broadcasting server for local training, whereas the server periodically collects all trained parameters to perform a global average and then redistributes the improved model back to the clients. After adequate training and updating iterations, usually termed as communication rounds, between the server and its associated clients, the objective function is able to converge to the global optimal, and the convergence property of FL can be quantitatively demonstrated.

\subsection{Difference between Security and Privacy}
Although security and privacy are used interchangeably in the literature, it is important to highlight the difference between them. On one hand, security issues refer to unauthorized/malicious access, change or denial to data. Such attacks are usually launched by hackers with expert knowledge of the target system or network. Hence, the fundamental three goals of security are confidentiality, integrity, and availability.

On the other hand, privacy issues generally refer to unintentional disclosure of personal information, usually from open-access data. For example, from a side-by-side comparison of a vote registration dataset and an anonymous set of healthcare sensor records (e.g., no individual's name and ID), an attacker may be able to identify certain individuals and learn about their health conditions. This is because some quasi-identifiers such as gender, birth date, and zip code are the same in both datasets. As can be seen from the above example, privacy attacks only require common sense and involve no hacking activities. The fundamental reason of privacy issues is that a seemingly harmless open dataset may contain clues to individual's private information in real life. Hence, alternative goals such as anonymity, unlinkability, and unobservability have been proposed for privacy protection.

\subsection{Security and Privacy Protection for FL}
During the learning process there exists several privacy and security issues, and we can generally clarify the corresponding protection methods into three categories: privacy protection at the client side, privacy protection at the server side, and security protection for the FL.

\subsubsection{Privacy protection at the client side}
In FL, clients will upload their learning results including parameter values and weights to the server, but they may not trust the server since a curious server might have a look at the uploaded data to infer private information. To alleviate this concern, clients can employ some privacy-preservation technologies as follows:
\begin{itemize}
  \item Perturbation: The idea of perturbation is adding noise to the uploaded parameters by clients. This line of work often uses differential privacy \cite{dwork2006calibrating} to obscure certain sensitive attributes until the third party is not able to distinguish the individual, thereby making the data impossible to be restored so as to protect user privacy. In \cite{geyer2017differentially}, authors introduced a differential privacy approach to FL in order to add protection to client-side data. However, the root of these methods still require that data are transmitted elsewhere and they usually involve a trade-off between accuracy and privacy, which needs adjustments.
  \item Dummy: The concept of dummy method stems from the location privacy protection \cite{1647865}. Dummy model parameters along with the true one will be sent to the server from clients, which may hide client's contribution during training. Because of the aggregation processed at the server, the system performance can still be guaranteed. 
\end{itemize}
\subsubsection{Privacy protection at the server side}
After collecting updated parameters from clients, the server will perform a weighted average to these parameters according to data size. However, when the server broadcasts the aggregated parameters to clients for model synchronizing, this information may leak as there may exist eavesdroppers. Thus, protections at the server side are also of significance.
\begin{itemize}
  \item Aggregation: The key idea of aggregation is collecting data or model parameters from different clients on the server side. After aggregation, the adversaries or the un-trustful server cannot inspect client information according to this aggregated parameters. {\color{black}In addition, in some scenarios, sever has the liberty to select clients with high quality parameters or non-sensitive requirements.} However, the question on how to design an appropriate aggregation mechanism is still a challenging task for current FL.
  \item Secure Multi-Party Computation (SMC): The root of SMC is using encryption to make individual devices' updates uninspected by a server, instead of only revealing the sum after a sufficient number of updates \cite{rosulek2017improvements}. In details, SMC is a four-round interactive protocol optionally enabled during the reporting phase of a given communication round. In each protocol round, the server gathers messages from all devices, then uses the set of device messages to compute an independent response and return to each device. The third round constitutes a commit phase, during which devices upload cryptographically masked model updates to the server. Finally, there is a finalization phase that devices reveal sufficient cryptographic secrets to allow the server to unmask the aggregated model update.
\end{itemize}
\subsubsection{Security protection for FL framework}
As for the security of the whole FL framework, it mainly considers the model-stealing attacks. Specially, any participant in FL may introduce hidden backdoor functionality into the joint global model, e.g., to ensure that an image classifier assigns an attacker-chosen label to images with certain features, or that a word predictor completes certain sentences with an attacker. Consequently, there are also some protecting measures on the security design for FL.
\begin{itemize}
\item Homomorphic Encryption: Homomorphic encryption \cite{papernot2016semi} is adopted to protect user data through parameters exchange under encryption mechanism. That is the parameters are coded before uploading, and the public-private decoding keys are also need to transmit, which may cause extra communication cost.
\item Back-door Defender: Existing defenses against backdoor attacks are not effective as most of them require access to the training data. In addition, the FL system cannot ensure all clients are not malicious and has no visibility into what participants are doing locally, and prevents anyone from auditing participants' updates to the joint model.
\end{itemize}
\section{Challenges on Private and Secure FL}

In this section, we clarify four main issues in the private and secure FL system, and propose specific discussions on each issue.
\subsection{Convergence: An Issue Caused by Privacy Protection}
As pointed out in \cite{Sahu2018On}, the theoretical convergence guarantees have not been fully explored in the federated average learning, even although recent works can provide approximate convergence guarantee to some extent. However, these works usually assumed unrealistic scenarios, e.g., (i) the data is either shared across devices or distributed in an independent and identically distributed (\emph{i.i.d.}) manner, and (ii) all devices are involved in communication at each round.

If privacy protection is considered, the convergence of FL cannot be guaranteed for the current system setting. The main reason is that learning parameters will be in a non \emph{i.i.d.} manner if perturbation method is applied at the client side. Moreover, even if the convergence can be satisfied when appropriate measures are proposed, the learning performance should be properly characterized. Previous work in \cite{abadi2016deep} has shown that convergence can be guaranteed when artificial noises are added into a deep learning network, but the learning accuracy decreases around 40\% when solving a MNIST classification problem. As such, the following aspects need to be addressed:
\begin{itemize}
  \item Theoretical results should be provided about the convergence of privacy-preserving FL.
  \item Learning performances, i.e., learning accuracy, communication rounds and variations of loss functions, need to be investigated when privacy protection is considered.
  \item Privacy protection algorithm, both theoretically and empirically, should be devised. In addition, the tradeoff between the privacy level and the convergency speed also needs further investigation.
\end{itemize}

To explain this using a concrete example, let us consider the perturbation method described in Section II-A. If artificial noises are added at the client side, the aggregated noise power will influence the updated system parameters, and these parameters are not i.i.d.. Thus, the global weighted parameters at the server side may appear differences from the original one without noises. When the SGD is applied, the descent trend may change to a different or even an opposite direction if inappropriate noise is added. In this way, we cannot guarantee the convergence of the algorithm. In addition, even if the convergence is satisfied, the reduction in convergency speed, i.e., the communication rounds between clients and the server, and the learning performance, i.e., the classification accuracy, should be carefully quantified and analyzed.
\subsection{Data Poisoning: A Security Issue}
In FL, clients, who previously acted only as passive data providers, can now observe intermediate model states and might contribute arbitrary updates as part of the decentralized training process. This creates an opportunity for malicious clients to manipulate the training process with little restriction. In particular, adversaries posing as honest clients can send erroneous updates that maliciously influence the performances of the training model, a process that is known as model poisoning.

Traditional poisoning attacks compromise the training data to change the model's behavior at inference time. Researchers have considered the situation when one of members of a FL system maliciously attacks others by allowing a backdoor to be inserted to filch others' data \cite{8765347}. They showed that an adversarial participant can infer membership as well as properties associated with a subset of the training data. In addition, some malicious clients may update unreasonable parameters, which in turn harm the system performance. On the other hand, there exists possible eavesdroppers during server broadcasting the intermediate machine learning model states. Thus,
the data poisoning on the security issues can be summarized as follows:
\begin{itemize}
  \item How to measure the loss performance if any malicious clients produce data or model poisoning?
  \item How to recognize and prevent these poisoning behaviors from clients?
  \item How to improve the security level by preventing eavesdroppers during the communication?
\end{itemize}
\subsection{Scaling Up Issue: A Privacy and Security Issue}
It is straightforward to extend the current FL system into a large one, e.g., hundreds or thousands of clients, due to the availability of high-performance and low-price devices. However, this vast scale will bring out several practical issues: device availability that correlates with the local data distribution in complex ways (e.g., time zone dependency); unreliable device connectivity and interrupted execution; orchestration of lock-step execution across devices with varying availability; and limited device storage and compute resources. All these issues can be concluded as scaling up issues, and  the most important and urgent issue is what will happen if more UEs are able to participate in FL. Specifically, the following aspects need to be addressed:
\begin{itemize}
\item If more UEs participate in FL, it will lead to less communication rounds thanks to more computations in each round, which should be an obvious advantage.
\item If more UEs participate in FL, there will be less impact of data poison attack because it becomes difficult for an adversary to control a large number of UEs.
\item If more UEs participate in FL, will it provide better privacy protection? The intuition is that hiding a UE in a larger dataset is easier than doing the same in a smaller dataset.
\end{itemize}

In summary, it is unknown whether having more UEs is helpful to reduce the learning time or accuracy, and we will provide related experimental results in Section IV-C. In addition, a typical wireless scenario for scheme designing and performance investigation that multiple communication modes, i.e., LTE, WiFi, 5G, etc., exists in the uploading process. Resources allocation for these multiple modes needs to be optimized as most of works are not considering wireless transmission. In a wireless setting, the communication links between the server and the clients are uncertain and imperfect and this effect needs to be carefully studied in the design of the FL system, especially in the large scale one \cite{yang2019scheduling}.
\subsection{Model Aggregation: A Security Issue}
The aggregation is mainly processed at the server after collecting individuals' parameters, and updates the global model. This process is particularly important as it should absorb the advantages of the clients and determine the end of learning. If protection method is applied at the client side, such as the perturbation applied before collecting model parameters, the aggregation cannot be simply a conventional averaging process. The main reasons can be concluded as: (i) the noise power of perturbation is increasing along with the number of clients; (ii) the server should know the stochastic information from clients and the design of the aggregation method needs to distinguish the privacy-sensitive clients from privacy-insensitive ones. Therefore, a more intelligent aggregation process should be provided as follows:
\begin{itemize}
  \item An intelligent aggregator should recognize the differences of clients and employ different aggregating strategies for them.
  \item An intelligent aggregator should resolve the noise-added problem provided by the privacy protection. For example, the use of minimum mean square estimation (MMSE) aggregator can serve as an effective candidate.
  \item An intelligent aggregator should update parameter weights for the participating clients during different communication rounds.
\end{itemize}

In particular, some form of recognition mechanism can be integrated into the aggregation process. It is able to adjust the parameter weights according to the quality of parameters or system feedback. Furthermore, some anomaly detection schemes can be considered to identify outliers during communications. The aggregator should be sufficiently intelligent as it can select appropriate clients for learning to achieve fast convergence and high performance.
\section{Experiment Results and Possible Solutions}
In this section, we provide simulations to demonstrate the aforementioned issues and discuss some possible solutions.
For each experiment, we first partition the original training data into disjoint non i.i.d. training sets, and locally compute SGD updates on each dataset, and then aggregate updates using an averaging method to train a globally shared classifier. We evaluate the prototype on the well-known classification dataset: MNIST, a digit classification problem which distinguishes 10 digital number from 0 to 9, and the system fails to complete the classification if the accuracy cannot exceed 10\%. The provided dataset in MNIST is divided into 60,000 training examples and 10,000 test examples. The global epoch is set to 300 iterations at the server side, while 120 iterations are implemented at each client side, and the local batch size is set to 1200. In the following figures, we collect 20 runs for each experiment and record the average results.

\subsection{Convergence}
In this subsection, we first show the classification accuracy with different noise powers, where local learning applies convolutional neutral network (CNN) system.
To achieve the privacy protection, we employ the perturbation method on the client side. In details, different artificial noises with same power i.e., gaussian noise $N_1\sim N(0,\delta)$ and Laplace noise $N_2\sim Lap(\lambda)$ are added to the local parameters, respectively.
\begin{figure}
\centering
 \includegraphics[width=0.4\textwidth]{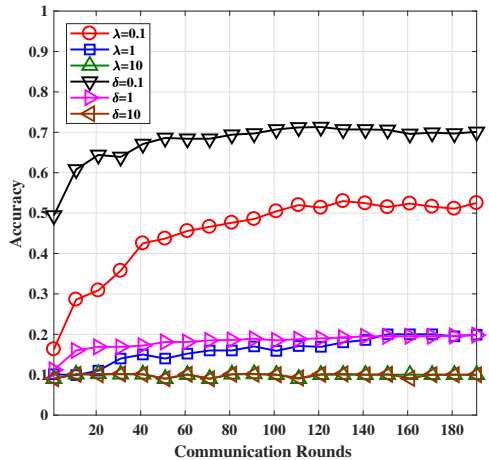}\\
\caption{Communication rounds versus accuracy with different noise powers in CNN}\label{noise-cnn}
\end{figure}
From the figure, we can observe that the accuracy performance is largely affected by the added noise while less influenced by the particular distribution of the noise.
In addition,
it will lead to poor performance or even system failure when large noises, i.e., $\lambda=10$, are added. This is due to the fact that the SGD algorithm has converged to a poor local minimum solution.

\begin{figure}
\centering
 \includegraphics[width=0.4\textwidth]{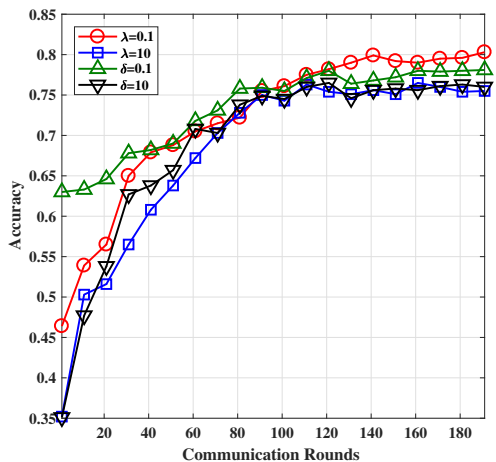}\\
\caption{Communication rounds versus accuracy with different noise powers in MLP}\label{noise-mlp}
\end{figure}
In addition, we verify this observation by applying multi-layer perception (MLP) system at clients. As can be seen in Fig.~\ref{noise-mlp}, the added noise seems to have slight influence on the accuracy. {\color{black} It it mainly because in the MLP system there is an auto-filtering process which can delete perceptions or parameters with bad performance.}

\textbf{Technical Problems:} The fundamental relationship between the convergence bound and the noise power needs to be characterized.

\textbf{Solution:} In the noise-added FL system, there is a fundamental tradeoff between the privacy level and learning performance. Intuitively speaking, a higher privacy level will incur more noise on the system, and lead to worse learning performance. Therefore, it is necessary to investigate the theoretical relationship between the noise scale, the local training iterations, the number of communication rounds as well as number of clients.
\subsection{Data Poisoning}

\begin{figure}
\centering
  \includegraphics[width=0.4\textwidth]{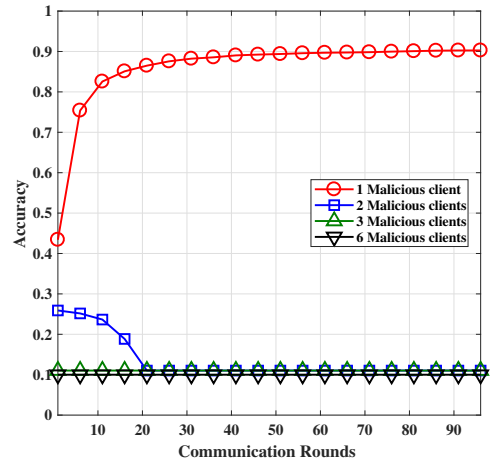}
  \caption{Performance comparison with different number of malicious clients in CNN}\label{poison}
\end{figure}
In Fig.~\ref{poison}, we show the performance comparison with different number of malicious clients. We set a CNN system for 30 clients, and the malicious clients will upload fake value of parameters in each communication round. The fake value can be the opposite of the true value, or random numbers within [-1, 1]. From Fig.~\ref{poison}, we can see that the system performance will be influenced if malicious clients exist. In addition, the system will fail when more malicious clients participating in.

\textbf{Technical Problems:} Mechanisms to prevent data poisoning need to be investigated.

\textbf{Solution:} There are three main ways to prevent the data poisoning in privacy-aware FL system. The first one is to recognize malicious clients when the system sets up. In this scenario, machine learning techniques can be utilized. For example, a supervised learning algorithm can be used to find malicious clients during each communication round. Another technique focuses on the aggregation process. After each aggregation, according to the quality of the uploaded learning parameters from the clients, the server can adjust the aggregation weights for each client. In this way, the server is able to put more confidence in the clients that are more helpful to achieve fast convergence and good learning performance. Third, concepts from social networks can be applied to update the weights in each communication round by exploiting the social relationship of each client to the overall system performance.
\subsection{Scaling Up Issue}
\begin{figure}
\centering
  \includegraphics[width=0.4\textwidth]{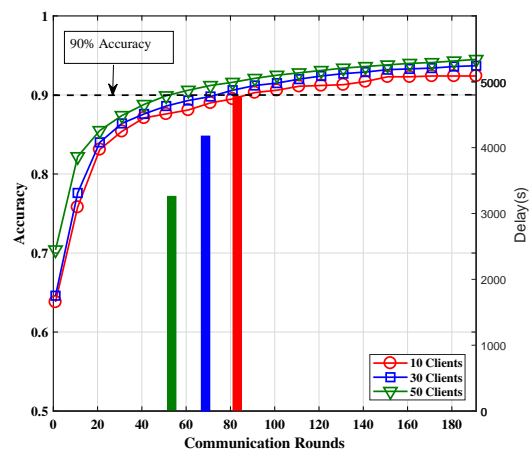}
  \caption{Performance comparison with different number of clients in CNN}\label{multi}
\end{figure}
In this subsection, we first show the classification accuracy with different clients numbers. From Fig.~\ref{multi} we can find that with the increasing number of clients, the performance does not show much gain. However, the total delay can be largely reduced when more clients exists. In particular, the clients are randomly distributed in a $1\times 1$ km$^2$ square area and we record the summation of the maximum calculation and transmission time as the delay in each communication round for different number of clients. Then we set the learning stops when the accuracy exceeds $90\%$ and record the total communication round, and calculate the total delay. Note that this result might be different for other delay models.

\textbf{Technical Problems:} In a large scale network, the server may suffer from a long waiting time and complicated resource allocation during parameter uploading.

\textbf{Solution:} For the scaling up issue, one promising method to address the long waiting time is setting up an upload delay deadline for each client. At each learning epoch, server will collect at least required clients' parameters before executing next round of FL. If the waiting time exceeds this deadline, the current learning epoch is abandoned. In addition, to deal with the large number of clients, we can use the concept of user clustering in game theory. By partitioning clients into different clusters factitiously, each cluster of clients will compete with each other to complete the learning goal. The server will also provide benefits in return. In this new structure design, the large number of clients will be separated by their common interests, similar physical location or same uploading ways.
\subsection{Model Aggregation}
The model aggregation should be intelligent. It not only deals with the large amount of noise while guaranteeing the system performance, but also applies various aggregation methods for different clients. In the traditional federated learning setting, the current strategy of the aggregation weight depends to the training size, but a more intelligent aggregator should be designed for multiple objectives. In addition, the selection for the updated parameters can also be adjusted. For example, the server can choose the uploading ones with better channel or parameter qualities.

\begin{figure}
\centering
  \includegraphics[width=0.4\textwidth]{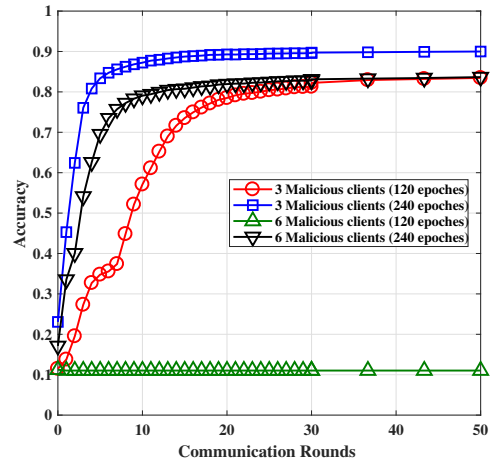}
  \caption{Performance comparison with different number of malicious clients under the proposed aggregation method in CNN}\label{aggregator}
\end{figure}

\textbf{Technical Problems:} We need to design an intelligent aggregator in the current FL setting.

\textbf{Solution:} In Fig.~\ref{aggregator}, we propose an intelligent aggregation method to address the malicious clients' problem. The proposed algorithm includes two parts: 1) Add a test process at the server side, and update the aggregation weight according to the testing performance for the uploaded parameters from each client. 2) Increase the local epochs for each client. As can be seen in the figure, the proposed algorithm can alleviate the performance degradation caused by the malicious clients. In addition, more local epochs are needed when more malicious clients exist in the FL system.

\section{Conclusion}
In this article, we have investigated potential privacy and security issues in federated learning (FL). We have pointed out that the privacy protection can be carried out at the client or the server side and security protection is mainly meant for the system level. In addition, we have argued that the considered issues can be classified into convergence, data poisoning, scaling up and model aggregation problems. Lastly, we have also provided some possible solutions for protecting privacy and security in designing FL systems.
\bibliographystyle{IEEEtran}
\bibliography{magzine}
\end{document}